\documentclass[graybox]{svmult}

\usepackage{mathptmx}       
\usepackage{helvet}         
\usepackage{courier}        
\usepackage{type1cm}        
\usepackage{makeidx}         
\usepackage{graphicx}        
\usepackage{multicol}        
\usepackage[bottom]{footmisc}

\usepackage{bm}
\usepackage{amsmath}
\usepackage{amssymb}
\usepackage{graphics}
\usepackage{fancybox}

\newcommand{\topline}{\hline\noalign{\smallskip}}
\newcommand{\midline}{\noalign{\smallskip}\svhline\noalign{\smallskip}}
\newcommand{\bottomline}{\noalign{\smallskip}\hline\noalign{\smallskip}}

\makeindex             

\usepackage{calligra}
\DeclareMathAlphabet{\mathcalligra}{T1}{calligra}{m}{n}
\DeclareFontShape{T1}{calligra}{m}{n}{<->s*[1.0]callig15}{}
\newcommand{\scriptp}{{\Large \calligra{p}}$\,$}
\begin{document}

\title*{Representation of Crystallographic Subperiodic 
Groups by Geometric Algebra}
\author{Eckhard Hitzer and Daisuke Ichikawa}
\institute{Eckhard Hitzer, Daisuke Ichikawa
\at Department of Applied Physics, 
University of Fukui, 910-8507 Fukui, Japan 
\email{hitzer@mech.fukui-u.ac.jp}, 
\email{ichikawa@hello.apphy.fukui-u.ac.jp}}
%
%
\maketitle

\abstract{We explain how following the representation of 3D 
crystallographic space groups in geometric algebra it is further possible to similarly 
represent the 162 socalled subperiodic groups of crystallography in 
geometric algebra. We construct a new compact geometric algebra 
group representation symbol, which allows to read off the complete set 
of geometric algebra generators. For clarity we moreover state 
explicitly what generators are chosen. The group symbols are based on 
the representation of point groups in geometric algebra by versors 
(Clifford group, Lipschitz elements). 
}

\abstract*{We explain how following the representation of 3D 
crystallographic space groups in geometric algebra it is further possible to similarly 
represent the 162 socalled subperiodic groups of crystallography in 
geometric algebra. We construct a new compact geometric algebra 
group representation symbol, which allows to read off the complete set 
of geometric algebra generators. For clarity we moreover state 
explicitly what generators are chosen. The group symbols are based on 
the representation of point groups in geometric algebra by versors 
(Clifford group, Lipschitz elements).}

\section{Introduction}

The 3D crystallographic space groups \cite{TH:ITA} have 
been successfully represented \cite{HH:CrystGA, DH:PGaSGinGA} in geometric algebra 
\cite{HLR:ConfM}. Following this an interactive 3D visualization has been 
created \cite{IH:OrthorSym, HP:TheSGV}. But for crystallographers the subperiodic 
space groups \cite{KL:ITE} in 2D and 3D with only one or two degrees 
of freedom for translations are also of great interest. 
\begin{figure}
\begin{center}
	\resizebox{0.7\textwidth}{!}{\includegraphics{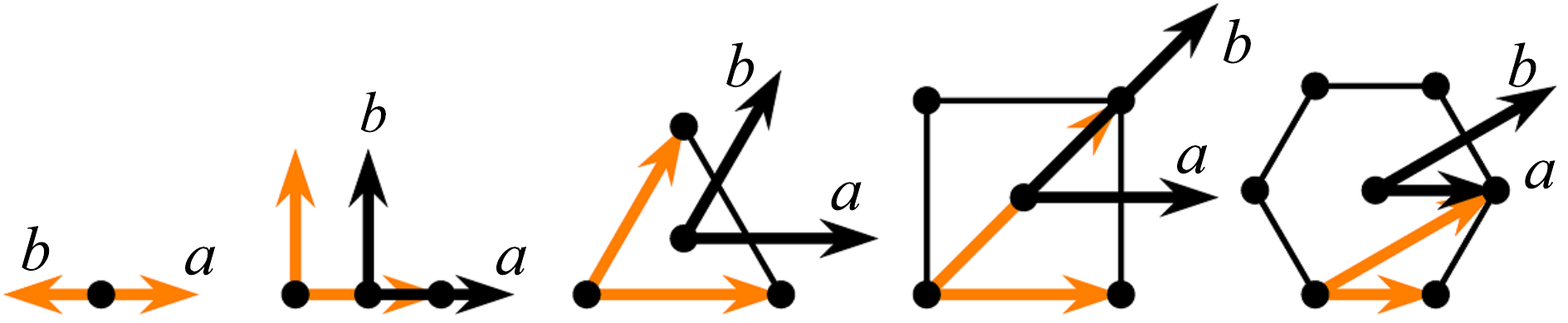}}
\caption{Regular polygons ($p=1,2,3,4,6$) and point 
group generating vectors $a,b$ subtending angles $\pi/p$ shifted to center.\label{fg:2Dpg}}
\end{center}
\end{figure}
\begin{figure}
 \begin{center}
	\resizebox{0.5\textwidth}{!}{\includegraphics{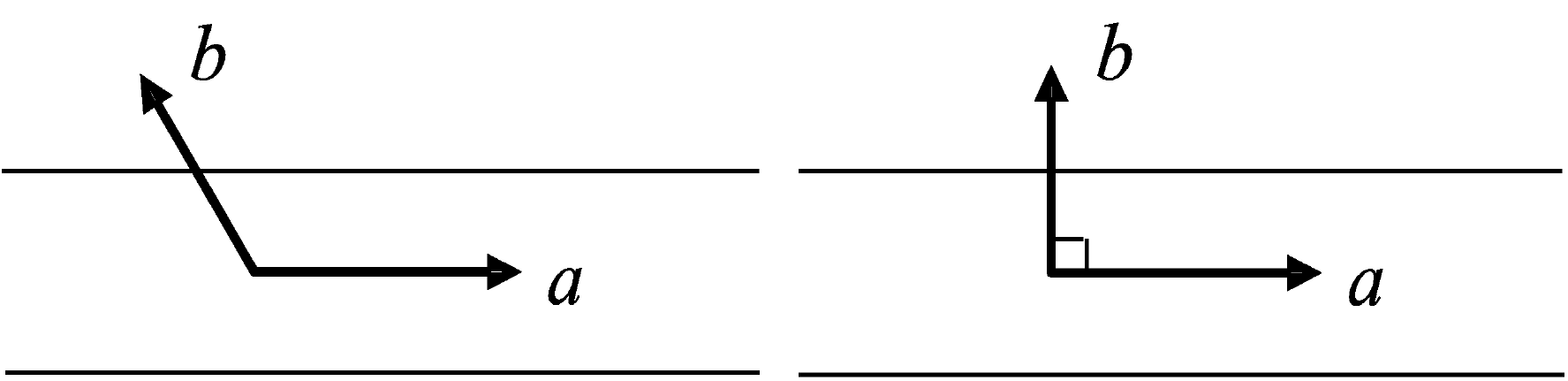}}
 \end{center}
\caption{Generating vectors $a,b$ of oblique and rectangular cells for 2D frieze groups.}
\label{fg:friezecells} 
\end{figure}
\begin{figure}
 \begin{center}  
	 \resizebox{0.9\textwidth}{!}{\includegraphics{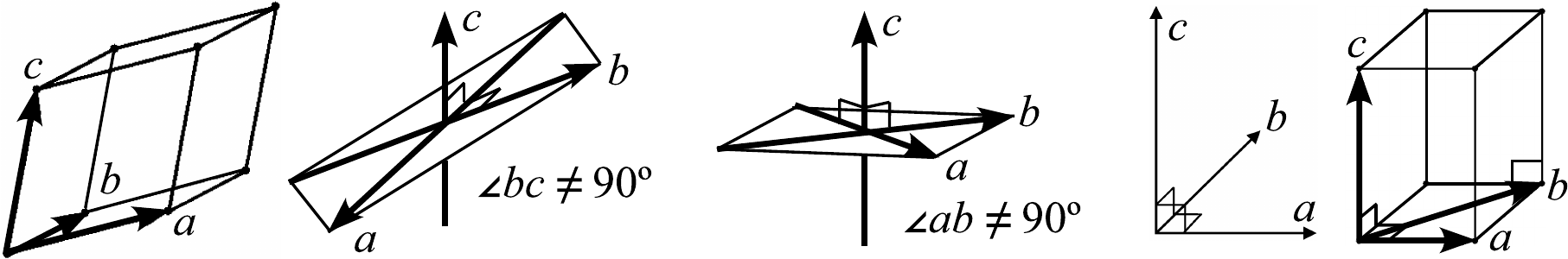}}
 \end{center}
\caption{From left to right: Triclinic, monoclinic inclined, monoclinic orthogonal, 
orthorhombic, and tetragonal cell vectors $a,b,c$ for rod and layer groups.}
\label{fg:rodcells}  
\end{figure}
\begin{figure}
 \begin{center}
  \resizebox{0.5\textwidth}{!}{\includegraphics{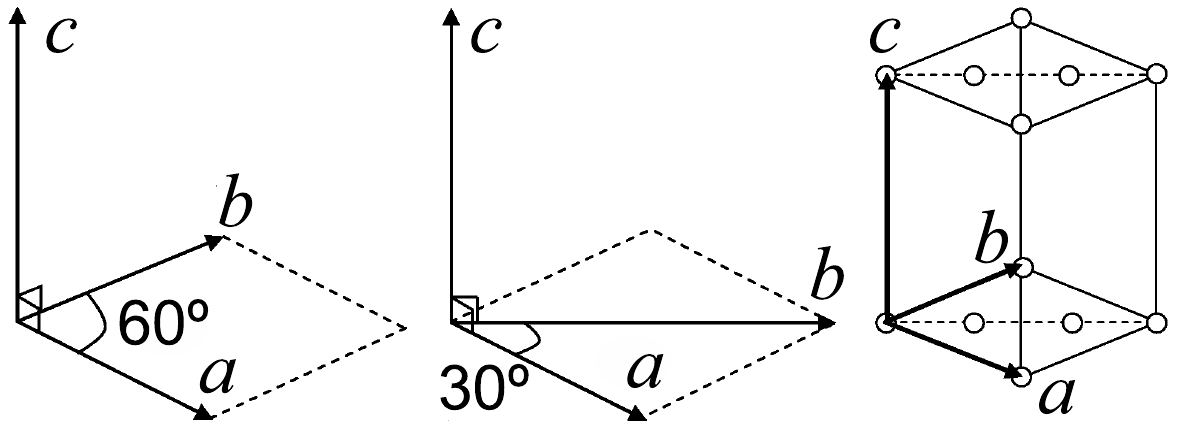}}
 \end{center}
\caption{Generating vectors $a,b,c$ of trigonal (left), hexagonal (center)
and hexagonally centered (right, Bravais symbol: $H$ or $h$)
cells for 3D rod and layer groups.}
\label{fg:trig-hex} 
\end{figure}
\begin{table}
\begin{center}
\caption{Geometric and international notation for 2D point groups.\label{tb:2Dpg}}
\begin{tabular}{lllllllllll}
\topline
Crystal&\multicolumn{2}{c}{Oblique}
&\multicolumn{2}{c}{Rectangular}
&\multicolumn{2}{c}{Trigonal}
&\multicolumn{2}{c}{Square}
&\multicolumn{2}{c}{Hexagonal}
\\
\midline
geometric & $\bar{1}$& $\bar{2}$ & 1& 2 & 3& $\bar{3}$ & 4&$\bar{4}$ & 6& $\bar{6}$
\\
international  &1& 2 &m& mm& 3m& 3& 4m& 4& 6m& 6\\
\bottomline
\end{tabular}
\end{center}
\end{table}
\begin{table}
\begin{center}
\caption{\label{tb:3Dpg}
Geometric 3D point group symbols~\cite{DH:PGaSGinGA} and generators with
${\theta_{a,b}=\pi/p}$,
${\theta_{b,c}=\pi/q}$,
${\theta_{a,c}=\pi/2}$, $p,q\in \{1,2,3,4,6\}$.}
\begin{tabular}{lcccccccc}
\topline
Symbol& $1$ & $p \neq 1$ & $\bar{p}$ & $pq$ 
& $\bar{p}q$ & $p\bar{q}$ & $\bar{p}\bar{q}$ & $\overline{pq}$
\\
\midline
Generators 
& $\,\,\, a \,\,\,$ 
& $\,\,\, a, \, b \,\,\, $
& $\,\,\,ab \,\,\,$ 
& $\,\,\,a, \, b, \, c \,\,\,$
& $\,\,\,ab, \, c \,\,\, $ 
& $\,\,\,a, \, bc \,\,\, $
& $\,\,\,ab, \, bc \,\,\, $ 
& $\,\,\,abc\,\,\,$
\\
\bottomline
\end{tabular}
\end{center}
\end{table}
\begin{table}
\begin{center}
\caption{ Table of frieze groups. 
Group number (col. 1),
intern. frieze group notation \cite{KL:ITE} (col. 2), 
related intern. 3D space group numbers \cite{TH:ITA} (col. 3),
and notation \cite{TH:ITA} (col. 4),
geometric 3D space group notation \cite{HH:CrystGA} (col. 5),
related intern. 2D space group numbers \cite{TH:ITA} (col. 6),
and notation \cite{TH:ITA} (col. 7),
related geometric 2D space group notation \cite{HH:CrystGA} (col. 8),
geometric frieze group notation (col. 9),
geometric algebra frieze group versor generators (col. 10).
The pure translation generator $T_a$ is omitted.}
\label{tb:frieze}
\begin{tabular}{cccccccccc}
\topline
 Frieze   & Intern. &  3D   & Intern. & Geom.  & 2D   & Intern. & Geom.  & Geom.  & Frieze Group \\
 Group \# & Notat.  &  SG\# & 3D SGN  & 3D SGN & SG\# & 2D SGN  & 2D SGN & Notat. & Generators \\  
\midline
\multicolumn{10}{l}{Oblique} \\ 
\midline
$F_{1}$ & \scriptp$1$   & 1 & $P1$ & $P\overline{1}$ & 1 & $p1$ & $p\overline{1}$ & \scriptp$\overline{1}$ &  \\ 
$F_{2}$ & \scriptp$211$ & 3 & $P2$ & $P\overline{2}$ & 2 & $p2$ & $p\overline{2}$ & \scriptp$\overline{2}$ & $a \wedge b$ \\
\bottomline
\multicolumn{10}{l}{Rectangular} \\ 
\midline
$F_{3}$ & \scriptp$1m1$ & 6  & $Pm$   & $P1$     & 3 & $pm(p1m1)$ & $p1$   & \scriptp1          & $a$ \\
$F_{4}$ & \scriptp$11m$ & 6  & $Pm$   & $P1$     & 3 & $pm(p11m)$ & $p1$   & \scriptp$\,.1$         & $b$ \\
$F_{5}$ & \scriptp$11g$ & 7  & $Pc$   & $P_{a}1$ & 4 & $pg(p11g)$ & $p_g1  $ & \scriptp${\,.\,}_{g}1$ & $b{T_{a}}^{1/2}$ \\
$F_{6}$ & \scriptp$2mm$ & 25 & $Pmm2$ & $P2$     & 6 & $p2mm$     & $p2$   & \scriptp2          & $a, b$ \\
$F_{7}$ & \scriptp$2mg$ & 28 & $Pma2$ & $P2_{a}$ & 7 & $p2mg$     & $p2_{g}$ & \scriptp$2_{g}$    & $a, b{T_{a}}^{1/2}$ \\
\bottomline
\end{tabular}
\end{center}
\end{table}
\begin{table}
\begin{center}
\caption{Table of triclinic, monoclinic and orthorhombic rod groups. 
The pure translator $T_c$ is omitted.}
\label{tb:rodgr1}
\begin{tabular}{ccccccc}
\topline
 Rod       & Intern. &  3D Space   & Intern. & Geom.  & Geom.  & Rod Group \\
 Group \#  & Notat.  &  Group \#   & 3D SGN  & 3D SGN & Notat. & Generators \\  
\midline
\multicolumn{7}{l}{Triclinic} \\ 
\midline
$R_{1}$ & \scriptp$1$ & $1$ & $P1$ & $P\bar{1}$ & \scriptp$\bar{1}$ &  \\ 
$R_{2}$ & \scriptp$2$ & $2$ & $P\bar{1}$ & $P\overline{22}$ & \scriptp$\overline{22}$ & $a \wedge b \wedge c$ \\
\bottomline
\multicolumn{7}{l}{Monoclinic/inclined} \\ 
\midline
$R_{3}$ & \scriptp$211$ & $3$ & $P112$ & $P\bar{2}$ &  \scriptp$\,.\,\bar{2}$ & $b \wedge c$ \\
$R_{4}$ & \scriptp$m11$ & $6$ & $Pm$ & $P1$ & \scriptp$1$ & $a$\\
$R_{5}$ & \scriptp$c11$ & $7$ & $Pc$ & $P_{c}1$ & \scriptp$_{c}1$ & $a{T_{c}^{1/2}}$ \\
$R_{6}$ & \scriptp$2/m11$ & $10$ & $P2/m$ & $P2\bar{2}$ & \scriptp$2\bar{2}$ & $a, b \wedge c$ \\
$R_{7}$ & \scriptp$2/c11$ & $13$ & $P2/c$ & $P_{a}2\bar{2}$ & \scriptp$_{c}2\bar{2}$ & $a{T_{c}^{1/2}}$, $b \wedge c$  \\
\bottomline
\multicolumn{7}{l}{Monoclinic/orthogonal} \\ 
\midline
$R_{8}$  & \scriptp$112$       & $3$  & $P112$      & $P\bar{2}$      & \scriptp$\bar{2}$     & $a \wedge b$ \\
$R_{9}$  & \scriptp$112_{1}$   & $4$  & $P2_{1}$    & $P\bar{2}_{1}$  & \scriptp$\bar{2}_{1}$ & $(a \wedge b){T_{c}^{1/2}}$ \\
$R_{10}$ & \scriptp$11m$       & $6$  & $Pm$        & $P1$            & \scriptp$\,..1$         & $c$ \\
$R_{11}$ & \scriptp$112/m$     & $10$ & $P2/m$      & $P\bar{2}2$     & \scriptp$\bar{2}2$    & $a \wedge b, c$ \\
$R_{12}$ & \scriptp$112_{1}/m$ & $11$ & $P2_{1}/m$  & $P\bar{2}_{1}2$ & \scriptp$\bar{2}_12$  & $(a \wedge b){T_{c}^{1/2}}, c$ \\
\bottomline
\multicolumn{7}{l}{Orthorhombic} \\ 
\midline
$R_{13}$ & \scriptp$222$ & $16$ & $P222$ & $P\bar{2}\bar{2}\bar{2}$ &  \scriptp$\bar{2}\bar{2}\bar{2}$ & $ab, bc$ \\ 
$R_{14}$ & \scriptp$222_{1}$ & $17$ & $P222_{1}$ & $P\bar{2}_{1}\bar{2}\bar{2}$ & \scriptp$\bar{2}_{1}\bar{2}\bar{2}$ & $ab{T_{c}^{1/2}}, bc$ \\
$R_{15}$ & \scriptp$mm2$ & $25$ & $Pmm2$ & $P2$ &  \scriptp$2$ & $a, b$ \\
$R_{16}$ & \scriptp$cc2$ & $27$ & $Pcc2$ & $P_{c}2_{c}$ &  \scriptp$_{c}2_{c}$ & $a{T_{c}^{1/2}}, b{T_{c}^{1/2}}$ \\
$R_{17}$ & \scriptp$mc2_{1}$ & $26$ & $Pmc2_{1}$ & $P2_{c}$ &  \scriptp$2_{c}$ & $a, b{T_{c}^{1/2}}$ \\
$R_{18}$ & \scriptp$2mm$ & $25$ & $Pmm2$ & $P2$ &  \scriptp$\,.2$ & $b, c$\\
$R_{19}$ & \scriptp$2cm$ & $28$ & $Pma2$ & $P2_{a}$ &  \scriptp$\,._{c}2$ & $b{T_{c}^{1/2}}, c$ \\
$R_{20}$ & \scriptp$mmm$ & $47$ & $Pmmm$ & $P22$ &  \scriptp$22$ & $a, b, c$ \\
$R_{21}$ & \scriptp$ccm$ & $49$ & $Pccm$ & $P_{c}2_{c}2$ &  \scriptp$_{c}2_{c}2$ & $a{T_{c}^{1/2}}, b{T_{c}^{1/2}}, c$ \\
$R_{22}$ & \scriptp$mcm$ & $51$ & $Pmma$ & $P22_{a}$ &  \scriptp$2_{c}2$ & $a, b{T_{c}^{1/2}}, c$ \\
\bottomline
\end{tabular}
\end{center}
\end{table}
\begin{table}
\begin{center}
\caption{Table of tetragonal and trigonal rod groups. 
The pure translator $T_c$ is omitted.}
\label{tb:rodg-tetra-trig}
\begin{tabular}{ccccccc}
\topline
 Rod       & Intern. &  3D Space   & Intern. & Geom.  & Geom.  & Rod Group \\
 Group \#  & Notat.  &  Group \#   & 3D SGN  & 3D SGN & Notat. & Generators \\  
\midline
\multicolumn{7}{l}{Tetragonal} \\ 
\midline
$R_{23}$ & \scriptp$4$ & $75$ & $P4$ & $P\bar{4}$ & \scriptp$\bar{4}$ & $ab$ \\ 
$R_{24}$ & \scriptp$4_{1}$ & $76$ & $P4_{1}$ & $P\bar{4}_{1}$ & \scriptp$\bar{4}_{1}$ & $ab{T_{c}}^{\frac{1}{4}}$ \\
$R_{25}$ & \scriptp$4_{2}$ & $77$ & $P4_{2}$ & $P\bar{4}_{2}$ &  \scriptp$\bar{4}_{2}$ & $ab{T_{c}}^{\frac{1}{2}}$ \\
$R_{26}$ & \scriptp$4_{3}$ & $78$ & $P4_{3}$ & $P\bar{4}_{3}$ & \scriptp$\bar{4}_{3}$ & $ab{T_{c}}^{\frac{3}{4}}$ \\
$R_{27}$ & \scriptp$\bar{4}$ & $81$ & $P\bar{4}$ & $P\overline{42}$ & \scriptp$\overline{42}$ & $abc$ \\
$R_{28}$ & \scriptp$4/m$ & $83$ & $P4/m$ & $P\bar{4}2$ & \scriptp$\bar{4}2$ & $ab$, $c$ \\
$R_{29}$ & \scriptp$4_{2}/m$ & $84$ & $P4_{2}/m$ & $P\bar{4}_{2}2$ & \scriptp$\bar{4}_{2}2$ & $ab{T_{c}}^{\frac{1}{2}}$, $c$ \\
$R_{30}$ & \scriptp$422$ & $89$ & $P422$ & $P\bar{4}\bar{2}\bar{2}$ & \scriptp$\bar{4}\bar{2}\bar{2}$ & $ab$, $bc$ \\
$R_{31}$ & \scriptp$4_{1}22$ & $91$ & $P4_{1}22$ & $P\bar{4}_{1}\bar{2}\bar{2}$ & \scriptp$\bar{4}_{1}\bar{2}\bar{2}$ & $ab{T_{c}}^{\frac{1}{4}}$, $bc$ \\
$R_{32}$ & \scriptp$4_{2}22$ & $93$ & $P4_{2}22$ & $P\bar{4}_{2}\bar{2}\bar{2}$ & \scriptp$\bar{4}_{2}\bar{2}\bar{2}$ & $ab{T_{c}}^{\frac{1}{2}}$, $bc$ \\
$R_{33}$ & \scriptp$4_{3}22$ & $95$ & $P4_{3}22$ & $P\bar{4}_{3}\bar{2}\bar{2}$ & \scriptp$\bar{4}_{3}\bar{2}\bar{2}$ & $ab{T_{c}}^{\frac{3}{4}}$, $bc$ \\
$R_{34}$ & \scriptp$4mm$ & $99$ & $P4mm$ & $P4$ & \scriptp$4$ & $a$, $b$ \\
$R_{35}$ & \scriptp$4_{2}cm$ & $101$ & $P4_{2}cm$ & $P_{c}4$ &  \scriptp$_{c}4$ & $a{T_{c}}^{\frac{1}{2}}$, $b$ \\ 
$R_{36}$ & \scriptp$4cc$ & $103$ & $P4cc$ & $P_{c}4_{c}$ & \scriptp$_{c}4_{c}$ & $a{T_{c}}^{\frac{1}{2}}$, $b{T_{c}}^{\frac{1}{2}}$ \\
$R_{37}$ & \scriptp$\bar{4}m2$ & $115$ & $P\bar{4}m2$ & $P4\bar{2}$ &  \scriptp$4\bar{2}$ & $a$, $bc$ \\
$R_{38}$ & \scriptp$\bar{4}c2$ & $116$ & $P\bar{4}c2$ & $P_{c}4\bar{2}$ &  \scriptp$_{c}4\bar{2}$ & $a{T_{c}}^{\frac{1}{2}}$, $bc$ \\
$R_{39}$ & \scriptp$4/mmm$ & $123$ & $P4/mmm$ & $P42$ &  \scriptp$42$ & $a$, $b$, $c$ \\
$R_{40}$ & \scriptp$4/mcc$ & $124$ & $P4/mcc$ & $P_{c}4_{c}2$ &  \scriptp$_{c}4_{c}2$ & $a{T_{c}}^{\frac{1}{2}}$, $b{T_{c}}^{\frac{1}{2}}$, $c$ \\
$R_{41}$ & \scriptp$4_{2}/mmc$ & $131$ & $P4_{2}/mmc$ & $P4_{c}2$ &  \scriptp$4_{c}2$ & $a$, $b{T_{c}}^{\frac{1}{2}}, c$ \\
\bottomline
\multicolumn{7}{l}{Trigonal} \\ 
\midline
$R_{42}$ & \scriptp$3$ & $143$ & $P3$ & $P\bar{3}$ & \scriptp$\bar{3}$ & $ab$ \\ 
$R_{43}$ & \scriptp$3_{1}$ & $144$ & $P3_{1}$ & $P\bar{3}_{1}$ & \scriptp$\bar{3}_{1}$ & $ab{T_{c}}^{\frac{1}{3}}$ \\
$R_{44}$ & \scriptp$3_{2}$ & $145$ & $P3_{2}$ & $P\bar{3}_{2}$ &  \scriptp$\bar{3}_{2}$ & $ab{T_{c}}^{\frac{2}{3}}$ \\
$R_{45}$ & \scriptp$\bar{3}$ & $147$ & $P\bar{3}$ & $P\overline{62}$ & \scriptp$\overline{62}$ & $abc$ \\
$R_{46}$ & \scriptp$312$ & $149$ & $P312$ & $P\bar{3}\bar{2}$ & \scriptp$\bar{3}\bar{2}$ & $ab$, $bc$ \\
$R_{47}$ & \scriptp$3_{1}12$ & $151$ & $P3_{1}12$ & $P\bar{3}_{1}\bar{2}$ & \scriptp$\bar{3}_{1}\bar{2}$ & $ab{T_{c}}^{\frac{1}{3}}$, $bc$ \\
$R_{48}$ & \scriptp$3_{2}12$ & $153$ & $P3_{2}12$ & $P\bar{3}_{2}\bar{2}$ & \scriptp$\bar{3}_{2}\bar{2}$ & $ab{T_{c}}^{\frac{2}{3}}$, $bc$ \\
$R_{49}$ & \scriptp$3m1$ & $156$ & $P3m1$ & $P3$ & \scriptp$3$ & $a$, $b$ \\
$R_{50}$ & \scriptp$3c1$ & $158$ & $P3c1$ & $P_{c}3_{c}$ & \scriptp$_{c}3_{c}$ & $a{T_{c}}^{\frac{1}{2}}$, $b{T_{c}}^{\frac{1}{2}}$ \\
$R_{51}$ & \scriptp$\bar{3}1m$ & $162$ & $P\bar{3}1m$ & $P\bar{2}6$     & \scriptp$6\bar{2}$       & $a$, $bc$ \\
$R_{52}$ & \scriptp$\bar{3}1c$ & $163$ & $P\bar{3}1c$ & $P\bar{2}_{c}6$ & \scriptp${_{c}}6\bar{2}$ & $a{T_{c}}^{\frac{1}{2}}$, $bc$ \\
\bottomline
\end{tabular}
\end{center}
\end{table}
\begin{table}
\begin{center}
\caption{Table of hexagonal rod groups. 
The pure translator $T_c$ is omitted.}
\label{tb:rodg-hex}
\begin{tabular}{ccccccc}
\topline
 Rod       & Intern. &  3D Space   & Intern. & Geom.  & Geom.  & Rod Group \\
 Group \#  & Notat.  &  Group \#   & 3D SGN  & 3D SGN & Notat. & Generators \\  
\midline
$R_{53}$ & \scriptp$6$ & $168$ & $P6$ & $P\bar{6}$ & \scriptp$\bar{6}$ & $ab$ \\ 
$R_{54}$ & \scriptp$6_{1}$ & $169$ & $P6_{1}$ & $P\bar{6}_{1}$ & \scriptp$\bar{6}_{1}$ & $ab{T_{c}}^{\frac{1}{6}}$ \\
$R_{55}$ & \scriptp$6_{2}$ & $171$ & $P6_{2}$ & $P\bar{6}_{2}$ & \scriptp$\bar{6}_{2}$ & $ab{T_{c}}^{\frac{1}{3}}$ \\
$R_{56}$ & \scriptp$6_{3}$ & $173$ & $P6_{3}$ & $P\bar{6}_{3}$ & \scriptp$\bar{6}_{3}$ & $ab{T_{c}}^{\frac{1}{2}}$ \\
$R_{57}$ & \scriptp$6_{4}$ & $172$ & $P6_{4}$ & $P\bar{6}_{4}$ & \scriptp$\bar{6}_{4}$ & $ab{T_{c}}^{\frac{2}{3}}$ \\
$R_{58}$ & \scriptp$6_{5}$ & $170$ & $P6_{5}$ & $P\bar{6}_{5}$ & \scriptp$\bar{6}_{5}$ & $ab{T_{c}}^{\frac{5}{6}}$ \\
$R_{59}$ & \scriptp$\bar{6}$ & $174$ & $P\bar{6}$ & $P\bar{3}2$ & \scriptp$\bar{3}2$ & $ab$, $c$ \\
$R_{60}$ & \scriptp$6/m$ & $175$ & $P6/m$ & $P\bar{6}2$ & \scriptp$\bar{6}2$ & $ab$, $c$ \\
$R_{61}$ & \scriptp$6_{3}/m$ & $176$ & $P6_{3}/m$ & $P\bar{6}_{3}2$ & \scriptp$\bar{6}_{3}2$ & $ab{T_{c}}^{\frac{1}{2}}$ \\
$R_{62}$ & \scriptp$622$ & $177$ & $P622$ & $P\bar{6}\bar{2}$ & \scriptp$\bar{6}\bar{2}$ & $ab$, $bc$ \\
$R_{63}$ & \scriptp$6_{1}22$ & $178$ & $P6_{1}22$ & $P\bar{6}_{1}\bar{2}$ & \scriptp$\bar{6}_{1}\bar{2}$ & $ab{T_{c}}^{\frac{1}{6}}$, $bc$ \\
$R_{64}$ & \scriptp$6_{2}22$ & $180$ & $P6_{2}22$ & $P\bar{6}_{2}\bar{2}$ & \scriptp$\bar{6}_{2}\bar{2}$ & $ab{T_{c}}^{\frac{1}{3}}$, $bc$ \\ 
$R_{65}$ & \scriptp$6_{3}22$ & $182$ & $P6_{3}22$ & $P\bar{6}_{3}\bar{2}$ & \scriptp$\bar{6}_{3}\bar{2}$ & $ab{T_{c}}^{\frac{1}{2}}$, $bc$ \\
$R_{66}$ & \scriptp$6_{4}22$ & $181$ & $P6_{4}22$ & $P\bar{6}_{4}\bar{2}$ & \scriptp$\bar{6}_{4}\bar{2}$ & $ab{T_{c}}^{\frac{2}{3}}$, $bc$ \\
$R_{67}$ & \scriptp$6_{5}22$ & $179$ & $P6_{5}22$ & $P\bar{6}_{5}\bar{2}$ & \scriptp$\bar{6}_{5}\bar{2}$ & $ab{T_{c}}^{\frac{5}{6}}$, $bc$ \\
$R_{68}$ & \scriptp$6mm$ & $183$ & $P6mm$ & $P6$ & \scriptp$6$ & $a$, $b$ \\
$R_{69}$ & \scriptp$6cc$ & $184$ & $P6cc$ & $P_{c}6_{c}$ & \scriptp$_{c}6_{c}$ & $a{T_{c}}^{\frac{1}{2}}$, $b{T_{c}}^{\frac{1}{2}}$ \\
$R_{70}$ & \scriptp$6_{3}cm$ & $185$ & $P6_{3}cm$ & $P_{c}6$ & \scriptp$_{c}6$ & $a{T_{c}}^{\frac{1}{2}}$, $b$ \\
$R_{71}$ & \scriptp$\bar{6}m2$ & $187$ & $P\bar{6}m2$ & $P32$ & \scriptp$32$ & $a$, $b$, $c$ \\
$R_{72}$ & \scriptp$\bar{6}c2$ & $188$ & $P\bar{6}c2$ & $P_{c}3_{c}2$ & \scriptp$_{c}3_{c}2$ & $a{T_{c}}^{\frac{1}{2}}$, $b{T_{c}}^{\frac{1}{2}}$, $c$ \\
$R_{73}$ & \scriptp$6/mmm$ & $191$ & $P6/mmm$ & $P62$ & \scriptp$62$ & $a$, $b$, $c$ \\
$R_{74}$ & \scriptp$6/mcc$ & $192$ & $P6/mcc$ & $P_{c}6_{c}2$ & \scriptp$_{c}6_{c}2$ & $a{T_{c}}^{\frac{1}{2}}$, $b{T_{c}}^{\frac{1}{2}}$, $c$ \\
$R_{75}$ & \scriptp$6_{3}/mcm$ & $193$ & $P6_{3}/mcm$ & $P_{c}62$ & \scriptp$_{c}62$ & $a{T_{c}}^{\frac{1}{2}}$, $b$, $c$ \\
\bottomline
\end{tabular}
\end{center}
\end{table}
\begin{table}
\begin{center}
\caption{Table of triclinic and monoclinic layer groups.
The pure translators $T_a, T_b$ are omitted.}
\label{tb:layer-tric-mon}
\begin{tabular}{ccccccc}
\topline
 Layer       & Intern. &  3D Space   & Intern. & Geom.  & Geom.  & Layer Group \\
 Group \#    & Notat.  &  Group \#   & 3D SGN  & 3D SGN & Notat. & Generators  \\  
\midline
\multicolumn{7}{l}{Triclinic/oblique} \\ 
\midline
$L_{1}$ & $p1$ & $1$ & $P1$ & $P\bar{1}$ & $p\bar{1}$ &  \\ 
$L_{2}$ & $p\bar{1}$ & $2$ & $P\bar{1}$ & $P\overline{22}$ & $p\overline{22}$ & $a \wedge b \wedge c$ \\ 
\bottomline
\multicolumn{7}{l}{Monoclinic/oblique} \\
\midline
$L_{3}$ & $p112$ & $3$ & $P2$ & $P\bar{2}$ &  $p\bar{2}$ & $a\wedge b$ \\ 
$L_{4}$ & $p11m$ & $6$ & $Pm$ & $P1$ &  $p..1$ & $c$ \\
$L_{5}$ & $p11a$ & $7$ & $Pc$ & $P_{a}1$ &  $p.._{a}1$ & $c{T_{a}}^{\frac{1}{2}}$ \\
$L_{6}$ & $p112/m$ & $10$ & $P2/m$ & $P\bar{2}2$ &  $p\bar{2}2$ & $a\wedge b$, $c$ \\
$L_{7}$ & $p112/a$ & $13$ & $P2/c$ & $P_{a}2\bar{2}$ &  $p\bar{2}2_{a}$ & $a\wedge b$, $c{T_{a}}^{\frac{1}{2}}$  \\
\bottomline
\multicolumn{7}{l}{Monoclinic/rectangular} \\
\midline
$L_{8}$ & $p211$ & $3$ & $P2$ & $P\bar{2}$ &  $p.\bar{2}$ & $b\wedge c$ \\
$L_{9}$ & $p2_{1}11$ & $4$ & $P2_{1}$ & $P\bar{2}_{1}$ &  $p.\bar{2}_{1}$ & $(b\wedge c){T_{a}}^{\frac{1}{2}}$\\
$L_{10}$ & $c211$ & $5$ & $C2$ & $A\bar{2}$ &  $c.\bar{2}$ & $b\wedge c$, ${T^{1/2}_{a+b}}$\\
$L_{11}$ & $pm11$ & $6$ & $Pm$ & $P1$ &  $p1$ & $a$\\
$L_{12}$ & $pb11$ & $7$ & $Pc$ & $P_{a}1$ &  $p_{b}1$ & $a{T_{b}}^{\frac{1}{2}}$\\
$L_{13}$ & $cm11$ & $8$ & $Cm$ & $A1$ &  $c1$ & $a$, ${T^{1/2}_{a+b}}$\\
$L_{14}$ & $p2/m11$ & $10$ & $P2/m$ & $P2\bar{2}$ &  $p2\bar{2}$ & $a$, $b\wedge c$\\
$L_{15}$ & $p2_{1}/m11$ & $11$ & $P2_{1}/m$ & $P2\bar{2}_{1}$ &  $p2\bar{2}_{1}$ & $a$, $(b\wedge c){T_{a}}^{\frac{1}{2}}$\\
$L_{16}$ & $p2/b11$ & $13$ & $P2/c$ & $P_{a}2\bar{2}$ &  $p_{b}2\bar{2}$ & $a{T_{b}}^{\frac{1}{2}}$, $b\wedge c$\\
$L_{17}$ & $p2_{1}/b11$ & $14$ & $P2_{1}/c$ & $P_{a}2\bar{2}_{2}$ &  $p_{b}2\bar{2}_{1}$ & $a{T_{b}}^{\frac{1}{2}}$, $(b\wedge c){T_{a}}^{\frac{1}{2}}$\\
$L_{18}$ & $c2/m11$ & $12$ & $C2/m$ & $A2\bar{2}$ &  $c2\bar{2}$ & $a$, $b\wedge c$, ${T^{1/2}_{a+b}}$\\
\bottomline
\end{tabular}
\end{center}
\end{table}
\begin{table}
\begin{center}
\caption{Table of orthorhombic/rectangular layer groups.
The pure translators $T_a, T_b$ are omitted.}
\label{tb:layer-orthorh}
\begin{tabular}{ccccccc}
\topline
 Layer       & Intern. &  3D Space   & Intern. & Geom.  & Geom.  & Layer Group \\
 Group \#    & Notat.  &  Group \#   & 3D SGN  & 3D SGN & Notat. & Generators  \\  
\midline
$L_{19}$ & $p222$ & $16$ & $P222$ & $P\bar{2}\bar{2}\bar{2}$ & $p\bar{2}\bar{2}\bar{2}$ & $ab$, $bc$ \\ 
$L_{20}$ & $p2_{1}22$ & $17$ & $P222_{1}$ & $P\bar{2}_{1}\bar{2}\bar{2}$ & $p\bar{2}\bar{2}_{1}\bar{2}$ & $ab$, $bc{T_{a}}^{\frac{1}{2}}$ \\ 
$L_{21}$ & $p2_{1}2_{1}2$ & $18$ & $P2_{1}2_{1}2$ & $P\bar{2}_{1}\bar{2}_{1}\bar{2}$ & $p\bar{2}\bar{2}_{1}\bar{2}_{1}$ & $bc{T_{a}}^{\frac{1}{2}}$, $ac{T_{b}}^{\frac{1}{2}}$ \\ 
$L_{22}$ & $c222$ & $21$ & $C222$ & $C\bar{2}\bar{2}\bar{2}$ & $c\bar{2}\bar{2}\bar{2}$ & $ab$, $bc$, ${T_{a+b}}^{\frac{1}{2}}$ \\ 
$L_{23}$ & $pmm2$ & $25$ & $Pmm2$ & $P2$ & $p2$ & $a$, $b$ \\ 
$L_{24}$ & $pma2$ & $28$ & $Pma2$ & $P2_{a}$ & $p2_{a}$ & $a$, $b{T_{a}}^{\frac{1}{2}}$ \\ 
$L_{25}$ & $pba2$ & $32$ & $Pba2$ & $P_{b}2_{a}$ & $p_{b}2_{a}$ & $a{T_{b}}^{\frac{1}{2}}$, $b{T_{a}}^{\frac{1}{2}}$ \\ 
$L_{26}$ & $cmm2$ & $35$ & $Cmm2$ & $C2$ & $c2$ & $a$, $b$, ${T^{1/2}_{a+b}}$ \\ 
$L_{27}$ & $pm2m$ & $25$ & $Pmm2$ & $P2$ & $p..2$ & $a$, $c$ \\ 
$L_{28}$ & $pm2_{1}b$ & $26$ & $Pmc2_{1}$ & $P2_{c}$ & $p.._{b}2$ & $a$, $c{T_{b}}^{\frac{1}{2}}$ \\ 
$L_{29}$ & $pb2_{1}m$ & $26$ & $Pmc2_{1}$ & $P2_{c}$ & $p_{b}..2$ & $a{T_{b}}^{\frac{1}{2}}$, $c$ \\ 
$L_{30}$ & $pb2b$ & $27$ & $Pcc2$ & $P_{c}2_{c}$ & $p_{b}.._{b}2$ & $a{T_{b}}^{\frac{1}{2}}$, $c{T_{b}}^{\frac{1}{2}}$ \\ 
$L_{31}$ & $pm2a$ & $28$ & $Pma2$ & $P2_{a}$ & $p.._{a}2$ & $a$, $c{T_{a}}^{\frac{1}{2}}$ \\ 
$L_{32}$ & $pm2_{1}n$ & $31$ & $Pmn2_{1}$ & $P2_{n}$ & $p.._{n}2$ & $a$, $c{T^{1/2}_{a+b}}$ \\ 
$L_{33}$ & $pb2_{1}a$ & $29$ & $Pca2_{1}$ & $P_{c}2_{a}$ & $p_{b}.._{a}2$ & $a{T_{b}}^{\frac{1}{2}}$, $c{T_{a}}^{\frac{1}{2}}$ \\ 
$L_{34}$ & $pb2n$ & $30$ & $Pnc2$ & $P_{n}2_{c}$ & $p_{b}.._{n}2$ & $a{T_{b}}^{\frac{1}{2}}$, $c{T^{1/2}_{a+b}}$ \\ 
$L_{35}$ & $cm2m$ & $35$ & $Cmm2$ & $C2$ & $c..2$ & $a$, $c$, ${T^{1/2}_{a+b}}$ \\ 
$L_{36}$ & $cm2e$ & $39$ & $Aem2$ & $A_{b}2$ & $c.._{a}2$ & $a$, $c{T_{a}}^{\frac{1}{2}}$, ${T^{1/2}_{a+b}}$ \\ 
$L_{37}$ & $pmmm$ & $47$ & $Pmmm$ & $P22$ & $p22$ & $a$, $b$, $c$ \\ 
$L_{38}$ & $pmaa$ & $49$ & $Pccm$ & $P_{c}2_{c}2$ & $p2_{a}2_{a}$ & $a$, $b{T_{a}}^{\frac{1}{2}}$, $c{T_{a}}^{\frac{1}{2}}$ \\ 
$L_{39}$ & $pban$ & $50$ & $Pban$ & $P_{b}2_{a}2_{n}$ & $p_{b}2_{a}2_{n}$ & $a{T_{b}}^{\frac{1}{2}}$, $b{T^{1/2}_{a}}$, $c{T_{a+b}}^{1/2}$ \\ 
$L_{40}$ & $pmam$ & $51$ & $Pmma$ & $P22_{a}$ & $p2_{a}2$ & $a$, $b{T_{a}}^{\frac{1}{2}}$, $c$ \\ 
$L_{41}$ & $pmma$ & $51$ & $Pmma$ & $P22_{a}$ & $p22_{a}$ & $a$, $b$, $c{T_{a}}^{\frac{1}{2}}$ \\ 
$L_{42}$ & $pman$ & $53$ & $Pmna$ & $P2_{n}2_{a}$ & $p2_{a}2_{n}$ & $a$, $b{T_{a}}^{\frac{1}{2}}$, $c{T^{1/2}_{a+b}}$ \\ 
$L_{43}$ & $pbaa$ & $54$ & $Pcca$ & $P_{c}2_{c}2_{a}$ & $p_{b}2_{a}2_{a}$ & $a{T_{b}}^{\frac{1}{2}}$, $b{T_{a}}^{\frac{1}{2}}$, $c{T_{a}}^{\frac{1}{2}}$ \\ 
$L_{44}$ & $pbam$ & $55$ & $Pbam$ & $P_{b}2_{a}2$ & $p_{b}2_{a}2$ & $a{T_{b}}^{\frac{1}{2}}$, $b{T_{a}}^{\frac{1}{2}}$, $c$ \\ 
$L_{45}$ & $pbma$ & $57$ & $Pbcm$ & $P_{b}2_{c}2$ & $p_{b}22_{a}$ & $a{T_{b}}^{\frac{1}{2}}$, $b$, $c{T^{1/2}_{a}}$ \\ 
$L_{46}$ & $pmmn$ & $59$ & $Pmmn$ & $P22_{n}$ & $p22_{n}$ & $a$, $b$, $c{T_{a+b}}^{\frac{1}{2}}$ \\ 
$L_{47}$ & $cmmm$ & $65$ & $Cmmm$ & $C22$ & $c22$ & $a$, $b$, $c$, ${T^{1/2}_{a+b}}$ \\ 
$L_{48}$ & $cmme$ & $67$ & $Cmme$ & $C22_{a}$ & $c22_{a}$ & $a$, $b$, $c{T_{a}}^{\frac{1}{2}}$, ${T^{1/2}_{a+b}}$ \\ 
\bottomline
\end{tabular}
\end{center}
\end{table}
\begin{table}
\begin{center}
\caption{Table of tetragonal, trigonal and hexagonal layer groups. 
$L_{57}$, $L_{58}$ and $L_{71}$ use special vector notation.
The pure translators $T_a, T_b$ are omitted.}
\label{tb:layer-tetra-tri-hex}
\begin{tabular}{ccccccc}
\topline
 Layer       & Intern. &  3D Space   & Intern. & Geom.  & Geom.  & Layer Group \\
 Group \#    & Notat.  &  Group \#   & 3D SGN  & 3D SGN & Notat. & Generators  \\  
\midline
\multicolumn{7}{l}{Tetragonal/square} \\ 
\midline
$L_{49}$ & $p4$ & $75$ & $P4$ & $P\bar{4}$ & $p\bar{4}$ & $ab$ \\ 
$L_{50}$ & $p\bar{4}$ & $81$ & $P\bar{4}$ & $P\overline{42}$ & $p\overline{42}$ & $abc$ \\ 
$L_{51}$ & $p4/m$ & $83$ & $P4/m$ & $P\bar{4}2$ & $p\bar{4}2$ & $ab$, $c$ \\
$L_{52}$ & $p4/n$ & $85$ & $P4/n$ & $P\bar{4}_{n}2$ & $p\bar{4}_{n}2$ & $ab$, $c{T_{b}}^{\frac{1}{2}}$ \\
$L_{53}$ & $p422$ & $89$ & $P422$ & $P\bar{4}\bar{2}\bar{2}$ & $p\bar{4}\bar{2}\bar{2}$ & $ab$, $bc$ \\
$L_{54}$ & $p42_{1}2$ & $90$ & $P42_{1}2$ & $P\bar{4}\bar{2}_{1}\bar{2}$ & $p\bar{4}\bar{2}_{1}\bar{2}$ & $ab$, $bc{T^{1/2}_{2a-b}}$ \\
$L_{55}$ & $p4mm$ & $99$ & $P4mm$ & $P4$ & $p4$ & $a$, $b$ \\
$L_{56}$ & $p4bm$ & $100$ & $P4bm$ & $P_{b}4$ & $p_{b}4$ & $a{T^{1/2}_{a-b}}$, $b$ \\
$L_{57}$ & $p\bar{4}2m$ & $111$ & $P\bar{4}2m$ & $P\bar{2}4$ & $p\bar{2}4$ & $ac$, $b$ \\
$L_{58}$ & $p\bar{4}2_{1}m$ & $113$ & $P\bar{4}2_{1}m$ & $P\bar{2}_{1}4$ & $p\bar{2}_{1}4$ & $ac{T^{1/2}_{a-b}}$, $b$ \\
$L_{59}$ & $p\bar{4}m2$ & $115$ & $P\bar{4}m2$ & $P4\bar{2}$ & $p4\bar{2}$ & $a$, $bc$ \\
$L_{60}$ & $p\bar{4}b2$ & $117$ & $P\bar{4}b2$ & $P_{b}4\bar{2}$ & $p_{b}4\bar{2}$ & $a{T^{1/2}_{a-b}}$, $bc$ \\
$L_{61}$ & $p4/mmm$ & $123$ & $P4/mmm$ & $P42$ &  $p42$ & $a$, $b$, $c$ \\
$L_{62}$ & $p4/nbm$ & $125$ & $P4/nbm$ & $P_{b}42_{n}$ &  $p_{b}42_{n}$ & $a{T^{1/2}_{a-b}}$, $b$, $c{T_{b}}^{\frac{1}{2}}$ \\
$L_{63}$ & $p4/mbm$ & $127$ & $P4/mbm$ & $P_{b}42$ &  $p_{b}42$ & $a{T^{1/2}_{a-b}}$, $b$, $c$ \\
$L_{64}$ & $p4/nmm$ & $129$ & $P4/nmm$ & $P42_{n}$ &  $p42_{n}$ & $a$, $b$, $c{T_{b}}^{\frac{1}{2}}$ \\
\bottomline
\multicolumn{7}{l}{Trigonal/hexagonal} \\
\midline
$L_{65}$ & $p3$ & $143$ & $P3$ & $P\bar{3}$ & $p\bar{3}$ & $ab$ \\ 
$L_{66}$ & $p\bar{3}$ & $147$ & $P\bar{3}$ & $P\overline{62}$ & $p\overline{62}$ & $abc$ \\
$L_{67}$ & $p312$ & $149$ & $P312$ & $P\bar{3}\bar{2}$ & $p\bar{3}\bar{2}$ & $ab$, $bc$ \\
$L_{68}$ & $p321$ & $150$ & $P321$ & $H\bar{3}\bar{2}$ & $h\bar{3}\bar{2}$ & $ab$, $bc$ \\
$L_{69}$ & $p3m1$ & $156$ & $P3m1$ & $P3$ & $p3$ & $a$, $b$ \\
$L_{70}$ & $p31m$ & $157$ & $P31m$ & $H3$ & $h3$ & $a$, $b$ \\
$L_{71}$ & $p\bar{3}1m$ & $162$ & $P\bar{3}1m$ & $P\bar{2}6$ & $p\bar{2}6$ & $ac$, $b$ \\
$L_{72}$ & $p\bar{3}m1$ & $164$ & $P\bar{3}m1$ & $P6\bar{2}$ & $p6\bar{2}$ & $a$, $bc$ \\
\bottomline
\multicolumn{7}{l}{Hexagonal/hexagonal} \\
\midline 
$L_{73}$ & $p6$ & $168$ & $P6$ & $P\bar{6}$ & $p\bar{6}$ & $ab$ \\ 
$L_{74}$ & $p\bar{6}$ & $174$ & $P\bar{6}$ & $P\bar{3}2$ & $p\bar{3}2$ & $ab$, $c$ \\
$L_{75}$ & $p6/m$ & $175$ & $P6/m$ & $P\bar{6}2$ & $p\bar{6}2$ & $ab$, $c$ \\
$L_{76}$ & $p622$ & $177$ & $P622$ & $P\bar{6}\bar{2}$ & $p\bar{6}\bar{2}$ & $ab$, $bc$ \\
$L_{77}$ & $p6mm$ & $183$ & $P6mm$ & $P6$ & $p6$ & $a$, $b$ \\
$L_{78}$ & $p\bar{6}m2$ & $187$ & $P\bar{6}m2$ & $P32$ & $p32$ & $a$, $b$, $c$ \\
$L_{79}$ & $p\bar{6}2m$ & $189$ & $P\bar{6}2m$ & $H32$ & $h32$ & $a$, $b$, $c$ \\
$L_{80}$ & $p6/mmm$ & $191$ & $P6/mmm$ & $P62$ & $p62$ & $a$, $b$, $c$ \\
\bottomline
\end{tabular}
\end{center}
\end{table}


\runinhead{Clifford geometric algebra.} 
Clifford's associative geometric product of two vectors simply adds 
the inner product to the outer product of Grassmann
\begin{equation}
  ab = a \cdot b + a \wedge b \,.
\end{equation}
This allows to write the reflection of a vector $x$ at a hyperplane 
through the origin with normal $a$ as
\begin{equation}
  x^{\,\,\prime}= -\,a^{\,\, -1} x\,a\, ,
  \quad \quad
   a^{\,\, -1} = \frac{a}{a^{\,\, 2}} \, .
\end{equation}
The composition of two reflections at hyperplanes whose normal vectors
$a,b$
subtend the angle $\alpha/2$ yields a rotation around the intersection of
the two hyperplanes by $\alpha$
\begin{equation}
  x^{\,\,\prime}
  = (ab)^{-1} x\,ab \, ,
  \quad \quad
  (ab)^{-1} = \,b^{\,\, -1}\,a^{\,\, -1} \, .
\end{equation}
In general the geometric product of $k$ normal vectors 
(the versor $S$) 
corresponds to the
composition of reflections to all symmetry transformations~\cite{DH:PGaSGinGA} of 
2D and 3D crystal cell point groups
\begin{equation}
 x^{\,\,\prime}
  = (-1)^k S^{\,-1} \,x \,S.
\end{equation}


\runinhead{Point groups.}
2D point groups are generated (cf. Table \ref{tb:2Dpg}) 
by multiplying vectors selected~\cite{DH:PGaSGinGA}  as in 
Fig. \ref{fg:2Dpg}. For example the hexagonal point group is given by multiplying its
two generating vectors $a,b$
\begin{equation}
 6 = \{
 a,b, R=ab, R^2, R^3, R^4, R^5, R^6=-1, 
 aR^2, bR^2, aR^4, bR^4
 \}.
\end{equation}
The rotation subgroups are denoted with bars, e.g. $\bar{6}$.
The selection of three vectors $a,b,c$ 
from each crystal cell~\cite{DH:PGaSGinGA,HP:TheSGV} 
for generating (cf. Table \ref{tb:3Dpg}) 3D
point groups are indicated in Figs. \ref{fg:rodcells} and \ref{fg:trig-hex}. 



\runinhead{Space groups.}
The smooth composition with translations is best done in the conformal
model~~\cite{HLR:ConfM} 
of Euclidean space (in the GA of $\mathbb{R}^{4,1}$). 
A plane can be described by the vector 
\begin{equation}
  m=p-d\,e_{\infty},
\end{equation}
where $p$ is a unit normal to the plane and $d$ its signed scalar distance 
from the origin. Reflecting at two parallel planes $m,m^{\prime}$ with 
distance $t/2$ we get the \textit{transla}tion opera\textit{tor} 
(by $t\in \mathbb{R}^3$)
\begin{equation}
  X^{\prime} = m^{\prime}m\,X\,mm^{\prime} = T_{\vec{t}}^{-1} X T_{t},
  \quad T_{t}=1+\frac{1}{2}te_{\infty}.
\end{equation}
Reflection at two non-parallel planes $m,m^{\prime}$ yields the rotation around
the $m,m^{\prime}$-intersection by twice the angle subtended by $m,m^{\prime}$.
Applying these techniques one can compactly tabulate geometric space group 
symbols and generators~\cite{DH:PGaSGinGA}, 
and consequently visualize them~\cite{HP:TheSGV} .

\section{Subperiodic groups represented in Clifford geometric algebra}

Compared to~\cite{HH:CrystGA} we have introduced dots: If one or two dots
occur between the Bravais symbol (\scriptp, $p$, $c$) and index $1$,
the vector $b$ or $c$, respectively, is present in the generator list. 
If one or two dots appear between the Bravais symbol and the index $2$
(without or with bar), then the vectors $b,c$ or $a,c$, respectively,
are present in the generator list.

In agreement~\cite{HH:CrystGA} the indexes $a,b,c,n$ (and $g$ for frieze groups) in first, 
second or third position 
after the Bravais symbol indicate that the reflections $a,b,c$ (in this order) become glide reflections. 
Index $n$ indicates
diagonal glides. The dots also serve as
position indicators. For example rod group $5$: \scriptp${_c1}$ has glide reflection
$aT_c^{1/2}$, rod group $19$: \scriptp${._c2}$ has $bT_c^{1/2}$, 
and layer group $39$: $p_b2_a2_n$ has $a{T_{b}}^{\frac{1}{2}}$, $b{T^{1/2}_{a}}$ 
and $c{T_{a+b}}^{1/2}$.

The notation $\overline{n}_p$ indicates a right handed screw rotation
of $2\pi/n$ around the $\overline{n}$-axis, with pitch $T^{p/n}_t$
where $t$ is the shortest lattice translation vector parallel to the axis, in
the screw direction. For example the layer group $21$: $p\bar{2}\bar{2}_1\bar{2}_1$
has the screw generators $bcT_a^{1/2}$ and $acT_b^{1/2}$.


\runinhead{Frieze groups.}
Figure \ref{fg:friezecells} shows the generating vectors $a, b$ of oblique and rectangular cells 
for 2D frieze groups. 
The only translation direction is $a$. Table \ref{tb:frieze} lists
the seven frieze groups with new geometric symbols and generators.


\runinhead{Rod groups.}
Figure \ref{fg:rodcells} shows the generating vectors $a, b, c$ of 
triclinic, monoclinic, orthorhombic and tetragonal cells 
for 3D rod and layer groups. 
Figure \ref{fg:trig-hex} shows the same for trigonal and hexagonal 
cells. 
For rod groups the only translation direction is $c$. 
Tables \ref{tb:rodgr1}, \ref{tb:rodg-tetra-trig}, \ref{tb:rodg-hex}  list
the 75 rod groups with new geometric symbols and generators:
Rod group number (col. 1),
intern. rod group notation \cite{KL:ITE} (col. 2), 
related intern. 3D space group numbers \cite{TH:ITA} (col. 3),
and notation \cite{TH:ITA} (col. 4),
related geometric 3D space group notation \cite{HH:CrystGA} (col. 5),
geometric rod group notation (col. 6),
geometric algebra generators (col. 7).

 
\runinhead{Layer groups.}
For layer groups the two translation directions are $a,b$. 
Tables \ref{tb:layer-tric-mon}, \ref{tb:layer-orthorh}, and \ref{tb:layer-tetra-tri-hex} list
the 80 3D layer groups with new geometric symbols and generators: 
Layer group number (col. 1),
intern. layer group notation \cite{KL:ITE} (col. 2), 
related intern. 3D space group numbers \cite{TH:ITA} (col. 3),
and notation \cite{TH:ITA} (col. 4),
related geometric 3D space group notation \cite{HH:CrystGA} (col. 5),
geometric layer group notation (col. 6),
geometric algebra generators (col. 7).
The layer groups are classified according to their 3D crystal system/2D Bravais system.
The monoclinic/oblique(rectangular) system corresponds to the
monoclinic/orthogonal(inclined) system of Fig. \ref{fg:rodcells}. 
Figure \ref{fg:trig-hex} shows the hexagonally centered cell 
with Bravais symbols $H$ (space group) and $h$ (layer group).

\begin{figure}
\begin{center}
  \resizebox{0.7\textwidth}{!}{\includegraphics{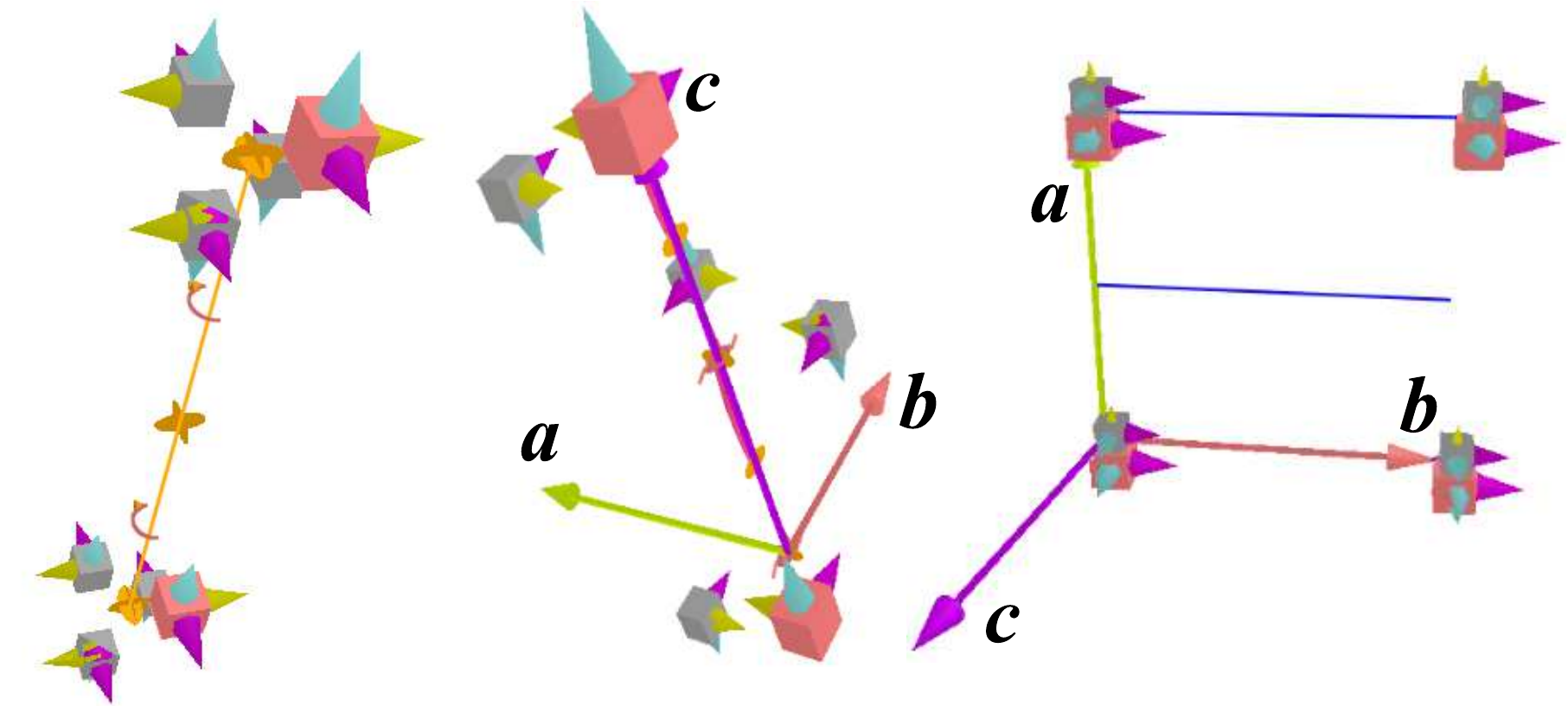}}
\caption{How a future subperiodic space group viewer software might depict
rod groups 13: \scriptp$\bar{2}\bar{2}\bar{2}$ 
and 14: \scriptp$\bar{2}_{1}\bar{2}\bar{2}$, 
and the layer group 11: $p1$, based on~\cite{HP:TheSGV}.\label{fg:extra}}
\end{center}
\end{figure}

\runinhead{Conclusion} 
We have devised a new Clifford geometric algebra representation
for the 162 subperiodic space groups using versors (Clifford group, Lipschitz elements). 
In the future this may be extended to magnetic subperiodic 
space groups, the sign of the generators may achieve that. 
We expect that the present work forms a suitable
foundation for interactive visualization software of subperiodic
space groups~\cite{HP:TheSGV}. 
Fig. \ref{fg:extra} shows how the 
rod groups 13: \scriptp$\bar{2}\bar{2}\bar{2}$ 
and 14: \scriptp$\bar{2}_{1}\bar{2}\bar{2}$, 
and the layer group 11: $p1$ and might be visualized in the future, 
based on~\cite{HP:TheSGV}. 

\begin{acknowledgement}
E. Hitzer wishes to thank God for his wonderful creation: 
\textit{You answer us with awesome deeds of righteousness,
       O God our Savior,
       the hope of all the ends of the earth
       and of the farthest seas ...}~\cite{Psalm65:5} 
He wishes to thank 
his family for their loving support, and 
D. Hestenes, C. Perwass, M. Aroyo, D. Litvin, A. Hayashi, N. Onoda, and Y. Koga.
\end{acknowledgement}

\end{document}